\documentclass[aps,prb,amsmath,amssymb,amsfont,showpacs,superscriptaddress,reprint]{revtex4-1}

\usepackage{amssymb,amsmath}
\usepackage{graphicx}
\usepackage{natbib}
\usepackage{multirow}
\usepackage{color}
\usepackage{graphicx}

\usepackage{lineno,hyperref}
\modulolinenumbers[5]

\begin{document}

\title{Doped carbon nanotubes as a model system of biased graphene}

\author{P. Szirmai\footnote{These authors contributed equally to this work.}}
\affiliation{Institute of Physics of Complex Matter, FBS Swiss Federal Institute of Technology (EPFL), CH-1015 Lausanne, Switzerland}

\author{B. G. M\'{a}rkus\footnote{These authors contributed equally to this work.}}
\affiliation{Department of Physics, Budapest University of Technology and Economics and MTA-BME Lend\"{u}let Spintronics Research Group (PROSPIN), P.O. Box 91, H-1521 Budapest, Hungary}

\author{B. D\'{o}ra}
\affiliation{Department of Theoretical Physics, Budapest University of Technology and Economics and MTA-BME Lend\"{u}let Exotic Quantum Phases Research Group (Momentum), P.O. Box 91, H-1521 Budapest, Hungary}

\author{G. F\'{a}bi\'{a}n}
\affiliation{Department of Physics, Budapest University of Technology and Economics and MTA-BME Lend\"{u}let Spintronics Research Group (PROSPIN), P.O. Box 91, H-1521 Budapest, Hungary}

\author{J. Koltai}
\affiliation{Department of Biological Physics, E\"{o}tv\"{o}s University, P\'{a}zm\'{a}ny P\'{e}ter s\'{e}t\'{a}ny 1/A, H-1117 Budapest, Hungary}

\author{V. Z\'{o}lyomi}
\affiliation{Physics Department, Lancaster University, Lancaster LA1 4YB, UK}

\author{J. K\"{u}rti}
\affiliation{Department of Biological Physics, E\"{o}tv\"{o}s University, P\'{a}zm\'{a}ny P\'{e}ter s\'{e}t\'{a}ny 1/A, H-1117 Budapest, Hungary}

\author{B. N\'{a}fr\'{a}di}
\affiliation{Institute of Physics of Complex Matter, FBS Swiss Federal Institute of Technology (EPFL), CH-1015 Lausanne, Switzerland}

\author{L. Forr\'{o}}
\affiliation{Institute of Physics of Complex Matter, FBS Swiss Federal Institute of Technology (EPFL), CH-1015 Lausanne, Switzerland}

\author{T. Pichler}
\affiliation{Faculty of Physics, University of Vienna, Strudlhofgasse 4., A-1090 Vienna, Austria}

\author{F. Simon}
\affiliation{Department of Physics, Budapest University of Technology and Economics and MTA-BME Lend\"{u}let Spintronics Research Group (PROSPIN), P.O. Box 91, H-1521 Budapest, Hungary}
\affiliation{Faculty of Physics, University of Vienna, Strudlhofgasse 4., A-1090 Vienna, Austria}

\begin{abstract}
Albeit difficult to access experimentally, the density of states (DOS) is a key parameter in solid state systems which governs several important phenomena including transport, magnetism, thermal, and thermoelectric properties. We study DOS in an ensemble of potassium intercalated single-wall carbon nanotubes (SWCNT) and show using electron spin resonance spectroscopy that a sizeable number of electron states are present, which gives rise to a Fermi-liquid behavior in this material. A comparison between theoretical and the experimental DOS indicates that it does not display significant correlation effects, even though the pristine nanotube material shows a Luttinger-liquid behavior. We argue that the carbon nanotube ensemble essentially maps out the whole Brillouin zone of graphene thus it acts as a model system of biased graphene. 
\end{abstract}


\maketitle

\section{Introduction}
There is a compelling link and similarity between the physical and chemical properties of the two carbon allotropes, graphene \cite{NovoselovSCI2004} and single-wall carbon nanotubes \cite{IijimaNAT1993,BethuneNAT1993}. Both contain carbon in a nearly sp$^2$ configuration and both consist essentially of a surface only. Concerning electronic properties, a linear energy dispersion is present for both materials but the differing dimensionality results in different energy dependence of the electronic density of states: for graphene, it is a smooth linear function of the energy \cite{GeimRMP}, whereas for SWCNTs, it contains Van Hove singularities \cite{HamadaPRL1992,MintmirePRL1992} whose presence is a fingerprint of the one-dimensional electronic character of SWCNTs. The similarity is even more striking for graphene nanoribbons (GNR) \cite{GNR_Louie_Nat2006} and SWCNTs \cite{HamadaPRL1992,MintmirePRL1992}: the electron confinement within the nanoribbon also gives rise to quantized states in a very similar manner to that of the SWCNTs. However, the quantization also means that SWCNTs and GNRs only map part of the graphene Brillouin-zone since only a subset of the graphene $k$-points are allowed for the two former materials.

In general, charge doping of graphite-based nanocarbon \cite{DresselhausAP2002} provides a way to yield insight into the electronic and vibrational properties as a function of the chemical potential (which is measured with respect to the charge neutral state). E.g. for graphite, a highly charge doped (or stage I) phase has the KC$_8$ stoichiometry for potassium doping, which involves a Fermi level shift of $1.35$ eV (Ref. \cite{GrueneisPRB2009a}). Compelling examples in graphene and SWCNTs, when charge doping led to interesting insights, include the emergence of intra-band transitions in graphene \cite{HeinzGrapheneOptics}, the bleaching of resonance Raman enhancement in SWCNTs \cite{RaoCNTRamanScience,EklundSWCNTDoping} and the Luttinger to Fermi liquid crossover in SWCNTs \cite{PichlerPRL2004}.

The electronic density of states, or DOS, is the central parameter for condensed matter systems \cite{AshcroftMermin}: it enters into most measurable properties such as e.g. electric or heat-transport and it governs strongly correlated phenomena such as, e.g., the superconducting transition \cite{BCS}. A comparison of experimental and calculated DOS values usually provide an elaborate way to test the accuracy of the theoretical description and whether strong correlation effects play a role.

Despite its importance, DOS is hardly accessible by direct means. Energy dependent DOS is measurable using photoemission and tunneling spectroscopy \cite{AshcroftMermin,STSReview}, however, both methods yield relative DOS values and the absolute value is accessible only upon extensive calibration. The value of the DOS at the Fermi level, $\text{DOS}(E_{\text{F}})$ is measurable by nuclear magnetic resonance spectroscopy \cite{SlichterBook} but this technique relies on the knowledge of the electron-nuclear hyperfine coupling constant. In contrast, the electronic specific heat and the Pauli spin-susceptibility of the conduction electrons provide a direct measurement of the DOS in the absence of strong correlation effects \cite{AshcroftMermin}. The Pauli spin-susceptibility is measurable by conduction electron spin resonance (CESR) experiments. CESR measures specifically the contribution of conduction electrons to magnetism \cite{WinterBook} and it was successful in, e.g., identifying strong correlation effects \cite{JanossyPRL1993} and the low-dimensional metallic character in fulleride conductors \cite{JanossyPRL1994}.

The following open questions called for a study of DOS in charge doped SWCNTs: \textit{i)} it is still debated whether strong correlation effects are present in chemically doped SWCNTs, \textit{ii)} it is also not known to what extent the SWCNT system can be used as a model system of biased graphene given the different dimensionality of their Fermi surfaces. Herein, we present conduction electron spin resonance studies on ensembles of SWCNTs with a well-defined diameter distribution under potassium doping to induce charge transfer to the tubes. We determine the DOS from the measurement of the Pauli spin-susceptibility of the conduction electrons and we compare this value with that obtained from tight-binding calculations. We find a good agreement between the two data, which confirms that an ensemble of chemically doped SWCNTs behaves as a Fermi-liquid with no substantial correlation effects. We show that the large number of different SWCNT geometries, which are present in our sample, give rise to a homogeneous mapping of almost \emph{all} $k$-points in the Brillouin zone of graphene.

\section{Experiments and calculations}

We used commercial SWCNTs prepared by the arc-discharge method from the same batch that we used previously for Raman measurements \cite{SimonPRB2005}, peapod filling \cite{SimonPRL2006}, and NMR studies \cite{SimonPRL2005}. According to Raman spectroscopy, the diameter distribution in the SWCNT samples is a Gaussian with a mean diameter of $d=1.4$ nm and variance of $\sigma=0.1$ nm. The material was purified with repeated air oxidation and acid treatments. In order to enable penetration of microwaves, thoroughly ground fine powder samples were prepared. Samples of about $5$ mg were vacuum annealed at $500^{\circ}\text{C}$ for 1 h in an ESR sample tube and inserted into an Ar filled glove-box without air exposure. We used two methods for the intercalation with potassium: the more conventional vapor technique \cite{DresselhausAP2002} and intercalation in liquid ammonia \cite{GalambosPSSB2009}. The vapor method involves sealing of the SWCNTs together with an abundant amount of potassium inside a quartz tube and an annealing to $200^{\circ}\text{C}$. This method works well for intercalating graphite and the surface of SWCNT samples but we found (as discussed below) that intercalation in ammonia is more efficient. Alkali metals are known to dissolve well in liquid ammonia which was used to synthesize alkali metal doped fullerides \cite{MurphyJPCS1992,RosseinskyAmmonia}, carbon nanotubes \cite{BillupsNL2004}, or graphene oxide \cite{AjayanACSNano2011}.

The saturated stoichiometry for K doping is around K:C$=1:7$ for SWCNTs \cite{PichlerPRL2001}, which is close to the KC$_8$ stoichiometry for graphite \cite{DresselhausAP2002}. To ensure saturation, about $30\%$ higher, non-stoichiometric amount of potassium was introduced into the ESR quartz tube which was subsequently placed under ammonia atmosphere. The quartz tube was inserted into an ethanol bath that was cooled down by liquid nitrogen. The doping hence proceeds at $-60^{\circ}$C in liquid ammonia promoted by slight sonication. The residual ammonia is evaporated by annealing at $200^{\circ}$C for $15$ minutes. The as-prepared material is inserted into a new, clean ESR quartz tube.

ESR measurements were performed using a commercial X-band spectrometer. $g$-factors were calibrated with respect to Mn$^{2+}$:MgO (the Mn$^{2+}$ content of MgO is $1.5$ ppm, and $g$-factor is $g=2.0014$ \cite{AbragamBook}) by taking into account the hyperfine interaction of Mn$^{2+}$ to second order. The ESR intensity was calibrated with CuSO$_4\times$5H$_2$O reference samples, to calibrate the spin-susceptibility. Raman spectrometry was carried out on a Labram (Horiba JY) spectrometer at $2.54$ eV ($488$ nm). Care was taken to avoid laser induced de-intercalation of the samples, the power was thus limited to $0.5$ mW. The spectral resolution was $2$ cm$^{-1}$.

We also performed microwave impedance measurements as a function of temperature with the so-called cavity perturbation method \cite{MehringRSI,KarsaPSSB}. This method yields the temperature-dependent resistivity (in relative units), which is otherwise unavailable for air-sensitive powder samples. The same samples were used for all studies.

The DOS was calculated in the nearest-neighbour tight-binding approximation for a large number ($81$) $(n,m)$ SWCNT chiralities as a function of the chemical potential \cite{HamadaPRL1992,MintmirePRL1998}. These data were then weighted with the abundance of each tubes, which was assumed to follow a Gaussian with the above mean diameter and variance \cite{Milnera2000,Jorio2001,Kuzmany2001,Borowiak2002,Liu2002,Saito2009,Mustonen2012,Chen2014}. Additional quasi-particle broadening of the Van Hove singularities of the SWCNTs due to finite life-time effects were considered \cite{SimonPRB2006}. The data for an $(n,m)$ SWCNT chirality appears along the so-called cutting lines due to the $k$-space quantization which corresponds to the circumference of the SWCNT \cite{HamadaPRL1992,MintmirePRL1998}. A smearing of the cutting lines due to the uncertainty principle was also included by broadening the cutting lines with a Gaussian function. 

\section{Results and discussion}

\begin{figure}[h!]
\begin{center}
\includegraphics[width=.65\linewidth]{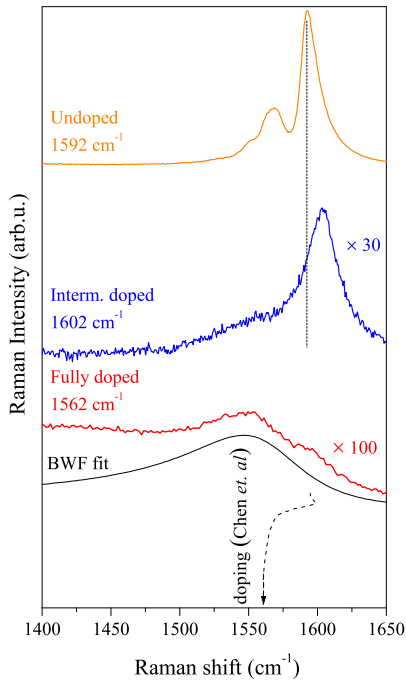}
\caption{G mode range Raman spectra of pristine, intermediately, and fully doped SWCNTs with a laser excitation of $\lambda=488$ nm. Upon intermediate doping, the $\mathrm{G}^-$ peak rapidly disappears and the $\mathrm{G}^+$ peak shows an upshift of $10$ cm$^{-1}$. For full doping, the $\mathrm{G}^+$ peak significantly downshifts and it has a Breit-Wigner-Fano lineshape (a fit is shown). Dashed curve shows how the $\mathrm{G}^+$ mode shifts upon doping according to Ref. \cite{EklundPRB2005} (the arrow points to increasing doping).}
\label{Raman_fig}
\end{center}
\end{figure}

In Fig. \ref{Raman_fig}., we show the G-mode range Raman spectra of pristine, intermediate doped, and fully potassium doped SWCNTs. The intermediate doping was achieved by the vapor phase method, whereas the full doping was performed by the liquid ammonia method. Characteristic changes are observed in the Raman spectra of the SWCNTs upon doping: the G mode component with lower Raman shift (known as $\mathrm{G}^-$ mode) broadens rapidly and vanishes \cite{Dresselhaus_doped_Gmode}. The G mode component with the higher Raman shift (known as the $\mathrm{G}^+$ mode) upshifts for the intermediate doping and significantly downshifts for the highest level of doping. Both observations agree well with the results of Raman studies on in-situ K and Cs doped SWCNTs in Ref. \cite{EklundPRB2005}, which proves that a saturated K intercalation is achieved in our samples for the liquid ammonia procedure. We find that the vapor doping does not produce homogeneously high doping levels for our relatively large sample amounts in contrast to doping thin sample films in the previous in-situ Raman study \cite{EklundPRB2005}.

\begin{figure}[h!]
\begin{center}
\includegraphics[width=.65\linewidth]{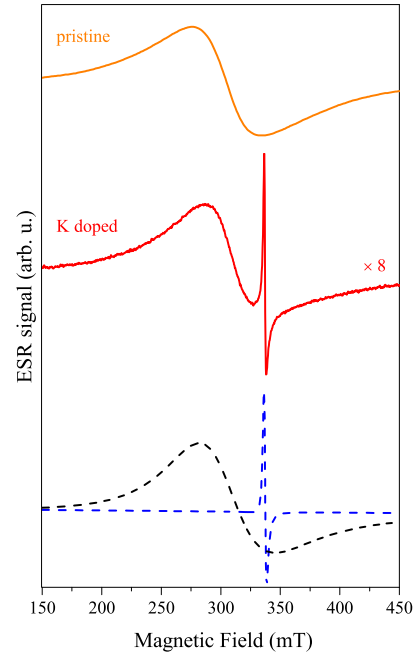}
\caption{ESR spectra of pristine and saturated K doped SWCNT samples at $T=300$ K. Note the broad background signal due to the catalyst particles, which is present already in the undoped sample and the narrow line which emerges upon doping. The deconvolution of the background and itinerant electron signal is shown for the doped sample as dashed curves.}
\label{ESR_fig}
\end{center}
\end{figure}

In Fig. \ref{ESR_fig}., we show the ESR spectra for the pristine and sample doped to saturation with potassium. An intensive and broad background due to the Ni:Y catalyst particles is observed for the pristine sample. The $g$-factor of this signal ($g_{\text{Ni}^{2+}} \approx 2.2$) allowed us to identify it as being due to the Ni$^{2+}$ ion with an intensity compatible with the expected Ni amount in the sample \cite{GalambosPSSB2009}. It was shown \cite{DoraPRL2008} that the undoped SWCNTs do not display the ESR signal of itinerant electrons due to the presence of the so-called Tomonaga-Luttinger liquid correlated state.

Upon doping, a narrower signal with an ESR line-width of $\Delta B_{\text{pp}}=2.2(1)$ mT emerges at $g=2.004(2)$ with an asymmetric lineshape. This lineshape is known as a Dysonian curve \cite{Dyson} and it can be fitted with a mixture of absorption and dispersion Lorentzian derivative lines \cite{WalmsleyJMR1996}. The microwave phase of the mixing is $37(5)^{\circ}$ that is close to the ideal $45^{\circ}$, which is expected when electrons with a low carrier mobility are embedded in a metal \cite{FeherKip}. We note that often the Dysonian, i.e., the asymmetric nature of an ESR signal is used as a hallmark that the ESR of itinerant electrons is observed. The present study shows that the signal due to Ni$^{2+}$ ions is \emph{also} asymmetric in the doped sample (see Fig. \ref{ESR_fig}). Any spin system whether due to localized or delocalized electrons (including nuclei), which is embedded in a metal gives an asymmetric lineshape \cite{FeherKip,SzirmaiPSSB2011}.

Simultaneously, the overall ESR signal intensity drops significantly due to the limited microwave penetration into the sample and due to a decrease in the cavity quality factor when the sample becomes more metallic. The narrower signal can be identified as being present due to the conduction electrons that are induced upon the charge transfer from the K to the SWCNTs. Several facts support this identification: the Pauli-like temperature dependence of the signal intensity as it is discussed below, the line-width matches well with that in K doped graphite powder, $2.2(1)$ mT, and follows a similar temperature dependence \cite{DresselhausAP2002}. We note that the presence of doped graphite powder can be excluded as the source of this signal as the graphite quantity in the pristine SWCNT samples is too small in the pristine SWCNT samples \cite{GalambosPSSB2009}.

\begin{figure}[h!]
\begin{center}
\includegraphics[width=.73\linewidth]{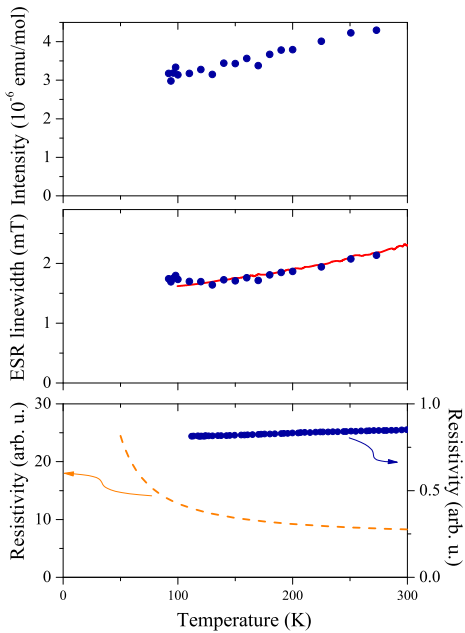}
\caption{Temperature dependence of the ESR intensity (a), linewidth (b), and sample resistivity (c) for the $100-300$ K range. The ESR linewidth data for K intercalated graphite (KC$_8$) from Ref. \cite{FabianSimon2012} is shown as a solid curve for comparison. The resistivity for the undoped sample is also shown for comparison in (c).}
\label{ESR_I_w_fig}
\end{center}
\end{figure}

In Fig. \ref{ESR_I_w_fig}., we show the temperature dependence of the ESR intensity and the linewidth of the ESR signal assigned to the conduction electrons. The temperature dependent sample resistivity is also shown. The data are not shown below $100$ K due to reasons discussed below. The ESR intensity shows a slight decrease with decreasing temperature (down to $100$ K), which proves that this signal indeed originates from the itinerant electrons. Were this signal coming from localized spins, its intensity would \emph{increase} by a factor $3$ when going from $300$ K to $100$ K. Indeed, the signal of the background grows in this manner (data not shown) and its intensity increases with $1/T$ as expected for localized spins from the Curie law. The slight decrease in the ESR intensity of itinerant electrons occurs due to a change in the microwave penetration into the sample and a change in the microwave cavity quality factor. Similar slight intensity drop was observed for metallic boron doped diamond \cite{SzirmaiBDD}, which was measured under identical conditions.

The ESR linewidth data in Fig. \ref{ESR_I_w_fig}b. agrees well for K doped SWCNTs and graphite, which also proves that the above identification is valid. This agreement also shows that the apparent difference in the structure of the two materials does not affect the ESR linewidth: for both materials $\Delta B_{\text{pp}}$ is dominated by spin scattering due to the K ions, which was explained for K doped graphite \cite{FabianSimon2012} in the framework of the Elliott-Yafet theory of spin-relaxation in metals \cite{Elliott,Yafet1963}. It is known from the study of alkali doped fullerides (A$_3$C$_{60}$, A$=$K and Rb) that for the heavier alkali elements (K and Rb) the ESR linewidth scales with the atomic spin-orbit coupling of the nuclei \cite{DoraSimonPRL2009}. However, a characteristic difference is expected for Li intercalated SWCNT and graphite as therein Li is expected to give a negligible contribution to the spin scattering.

Upon doping, the sample resistivity drops by about a factor $10$ at room temperature and its temperature dependence changes character from semiconducting to metallic behavior as shown in Fig. \ref{ESR_I_w_fig}c. These observations agree with previous studies \cite{FuhrerSSC1998,PichlerPRL2001} and prove that doping makes our powder samples metallic.

Our CESR data below $100$ K do not follow the trends described above, due the presence of paramagnetic impurities in our system. Similarly to the case of K$_3$C$_{60}$ \cite{NemesPRB2000}, the temperature-dependence of the ESR measurables approach the behavior of paramagnetic impurities. This indicates that at these temperatures the so-called \textit{bottleneck regime} is realized, i.e., a strong coupling is present between the paramagnetic and metallic spin systems.

In the following, we discuss the \emph{absolute} value of the static spin-susceptibility, $\chi_0$ and DOS in K doped SWCNTs. Due to its selection rules, ESR is selective to magnetism which originates from a spin quantum number, i.e., due to the Curie or Pauli susceptibility and is insensitive, e.g., to the Van Vleck or Landau susceptibilities \cite{SlichterBook}. For this reason, the ESR signal intensity is often mentioned to be a direct measure of the \emph{spin-susceptibility}. To obtain absolute values of $\chi_0$, a calibration of the ESR signal is required as it is detailed in Ref. \cite{SzirmaiBDD}. In brief, the ESR signal of a well known paramagnetic intensity standard (CuSO$_4\times$5H$_2$O in our case) is measured which allows to relate the ESR signal intensity to an actual spin-susceptibility. In principle it allows to determine $\chi_0$ for any samples.

For the present measurements, an intermediate calibration step is required as the K doping induces a change in the sample conductivity and thus the penetration of microwaves is also affected. The presence of this effect is clear from a signal intensity drop of about a factor $6(1)$ of the Ni$^{2+}$ ions. To take this effect into account, the measured signal intensity of the K doped SWCNTs is scaled back with the same factor. From our two-step calibration, we obtain a $\chi_0=4.3(9)\times 10^{-6}$ emu$/$mol. The sizeable error of this value arises from the somewhat uncertain amount of SWCNTs in the sample. The measured $\chi_0$ static spin-susceptiblity is related to the DOS through the Pauli susceptibility: 
\begin{equation}\label{Pauli}
\chi_0=\mu_0\frac{g^2}{4}\mu_{\text{B}}^2 \frac{\text{DOS} }{V_{\text{c}}},
\end{equation}
where $\mu_0$ is the vacuum permeability, $\mu_{\text{B}}$ is the Bohr magneton, $V_{\text{c}}$ is the volume of the unit cell \cite{AshcroftMermin}.

\begin{figure}[h!]
\begin{center}
\includegraphics[width=.9\linewidth]{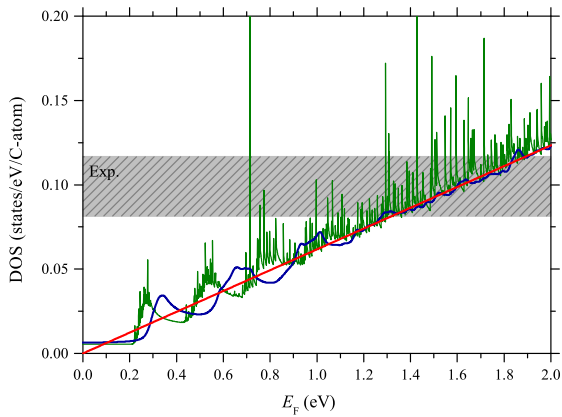}
\caption{Density of states as a function of chemical potential. The shady region indicates the experimentally determined DOS including error. Curves depict the DOS in graphene (red), in SWCNTs with the tight-binding approximation (TBA) at $T=0$ K (green), and TBA with room-temperature ($\sim 26$ meV) quasi-particle broadening (blue). The apparent shifting of the latter curve is a mathematical consequence of the convolution of the step-like DOS with the Lorentzian function.}
\label{DOS_fig}
\end{center}
\end{figure}

To test the validity of the experimentally determined DOS, we compare it with calculations that were performed on an ensemble of SWCNT as described above. Fig. \ref{DOS_fig} shows a comparison between the experimental DOS result and theoretical calculations as a function of the chemical potential, i.e., the energy separation from the Dirac point. The theoretical data is shown with zero and a finite $300$ K ($26$ meV) broadening parameter. For comparison, the DOS for graphene in the vicinity of the Dirac point is also shown according to Ref. \cite{GeimRMP}: $\text{DOS}(E)=A_{\text{c}}/\pi\times\left|E\right|/(\hbar v_{\text{F}})^2$, where $v_{\text{F}}=1.07\times10^6~\text{m}/\text{s}$ is the corresponding Fermi velocity \cite{AndreiPRB2009} and $A_{\text{c}}=5.24~\mathrm{\AA^2/unit~cell}$ is the area of the first Brillouin zone (BZ). The gray bar in Fig. \ref{DOS_fig} shows the value of the experimental DOS, which allows to deduce the chemical potential shift due to doping. The integration of a given DOS($E$) function yields the charge transfer. From this, we obtain that our experimental DOS corresponds to a stoichiometry of K:C=$1:(7\pm 1)$. Remarkably, the calculated K to C ratio is close to that found in alkali intercalated graphite (KC$_8$~\cite{DresselhausAP2002}) and in vapor-phase doped SWCNTs using \emph{in-situ} electron energy-loss spectroscopy \cite{Liu2003}. Within the limitations posed by the error bar, the experimental and theoretically deduced DOS data are in accordance. This finding means that the system under study does not show strong correlations. In view of the underlying one-dimensional character of the SWCNTs, this observation might sound surprising. However, several previous studies identified the intercalation-induced transition from the Tomonaga-Luttinger liquid to a three-dimensional Fermi-liquid phase \cite{Liu2003,PichlerPRL2004,Simon2008,GalambosPSSB2009}.

An ensemble of SWCNTs contains a large number of tubes with chiralities which follow a Gaussian diameter distribution \cite{SimonPRB2005,SimonPRL2005,Jorio2001,Kuzmany2001,Liu2002,Chen2014}. One expects that the eventual differences in the SWCNT geometries are smeared out for this ensemble, or even the one-dimensional characters are less pronounced and that it is possible to approach graphene, the mother compound of the SWCNTs.

\begin{figure}[h!]
\begin{center}
\includegraphics[width=0.83\linewidth]{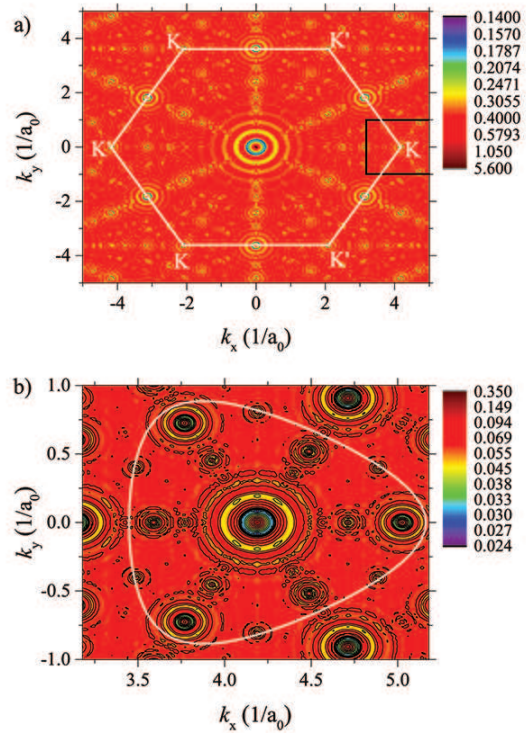}
\caption{Reciprocal space distribution of the cutting lines calculated for a typical nanotube ensemble for (a) the first BZ and (b) in the vicinity of the $\mathbf{K}$ point. The color map from blue to red illustrates the coverage of the possible electronic states. The equi-energetic contour line shown in white corresponds to KC$_7$ saturation doping ($E_{\text{F}}\approx 1.73$ eV). Note the C$_3$ symmetry around the $\mathbf{K}$ point.}
\label{clines_fig}
\end{center}
\end{figure}

To test this suggestion, we performed calculations investigating to what extent the BZ of graphene is mapped out by a carbon nanotube ensemble. In the zone-folding scheme, the one-dimensional representations of the quantized momentum-space directions of carbon nanotubes \cite{SamsonidzeAPL2004}, i.e., the cutting lines, display the electronic states of a given ($n,m$) chirality. In Fig. \ref{clines_fig}., we depict the reciprocal space coverage of a carbon nanotube ensemble in the proximity of the $\mathbf{K}$ point as the sum of the probability amplitudes of each of the cutting lines of all relevant chiralities. The projection illustrates that \emph{all} the electronic states of graphene are almost homogeneously represented by the carbon nanotube ensemble. The high-symmetry $\mathbf{K}$ point (as a result of the crossing of the metallic cutting lines) is slightly over-represented but the coverage at chemical shifts close to our doping (K:C$=1:7$) barely oscillates around its mean value \cite{SupMat}. Therefore, this illustration proves directly that at high doping levels the SWCNT ensemble behaves as the model system of doped or gate-biased graphene. This provides an additional link between graphene and carbon nanotubes: for an ensemble of SWCNTs the bulk properties (spin and thermal properties) mimic those in graphene and related compounds. In turn, such physical properties of a graphene could be studied using carbon nanotubes ensembles, in particular as a function of charge doping.

In conclusion, we found that using liquid ammonia doping, the KC$_7$ saturation potassium doping can be achieved in large quantities of SWCNT ensembles. We demonstrated that ESR is applicable to determine the electronic density of states in alkali doped SWCNTs, and we confirmed the absence of strong correlation effects, and that the material behaves as a three-dimensional Fermi liquid. By comparing the reciprocal space of a SWCNT ensemble and biased graphene, we illustrated that potassium-doped SWCNTs provide a tunable model system for graphene.

\section*{Acknowledgments}
Work supported by the ERC Grant Nr. ERC-259374-Sylo and the Swiss NSF (Grant No. 200021\_144419).

\section{Supplementary material}

\subsection{Low-temperature ESR measurements}

\begin{figure}[h!]
\begin{center}
\includegraphics[width=.83\linewidth]{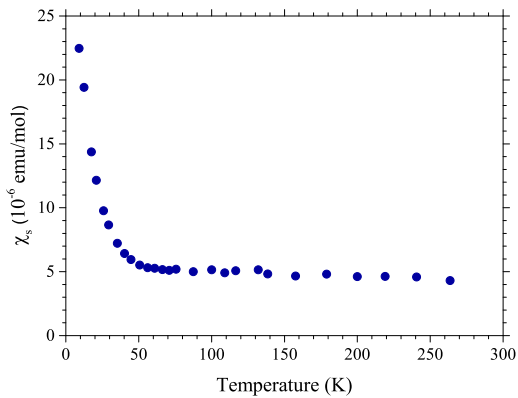}
\caption{Low-temperature spin-susceptibility of saturated potassium doped SWCNTs. Note that the low-temperature increase is provoked by the presence of a small amount of paramagnetic impurities.}
\label{SI_ESR_I_low_temp}
\end{center}
\end{figure}

Figure \ref{SI_ESR_I_low_temp}. depicts the spin-susceptibility at low temperatures, as determined from the ESR intensity. Below $100$ K, the intensity slightly increases indicating the presence of low amount of paramagnetic impurities ($0.6(1)\%$) in our system that give a single resonance as a result of the bottleneck effect \cite{BarnesAdvPhys}. The conditions of the bottleneck regime are quite naturally satisfied as the $g$-factor of the defects in carbon structures are expected to be close to the $g\approx 2$ CESR signal of potassium doped SWCNTs.

\begin{figure}[h!]
\begin{center}
\includegraphics[width=.83\linewidth]{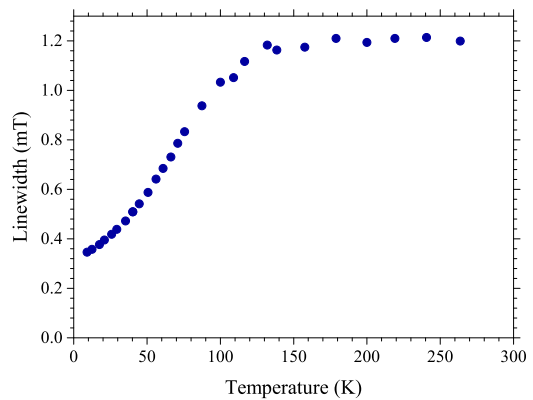}
\caption{Low-temperature ESR linewidth of K doped SWCNTs. The low-temperature narrowing is a result of the bottleneck effect.}
\label{SI_lw_low_temp}
\end{center}
\end{figure}

This assignment is further confirmed by the ESR linewidth (see Fig. \ref{SI_lw_low_temp}.), by the resonant field (not shown), and by the change of the phase of the mixture of the absorption and dispersion Lorentzian derivative lines. The change of the slope of the ESR linewidth around $100$ K indicates that the susceptibility of the impurities starts to dominate below this temperature. It thus causes the narrowing of the ESR line. Similarly, the $g$-factor slowly changes below $100$ K ($g(4~\text{K})-g(100~\text{K})=2.5(5)\times 10^{-4}$), highlighting that the $g$-factor of our ESR line is a mixture of the $g$-factor of the paramagnetic impurities and the $g$-factor of the CESR line.

\subsection{Raman measurements}

\begin{figure}[h!]
\begin{center}
\includegraphics[width=.83\linewidth]{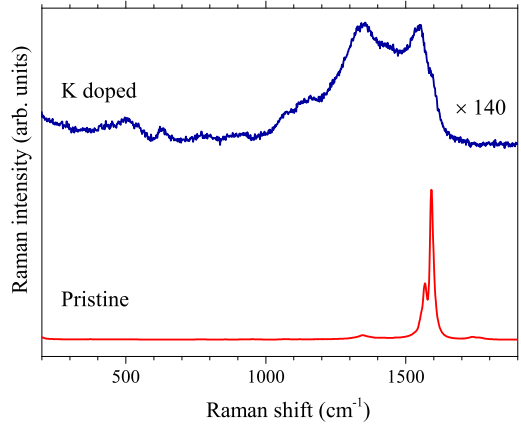}
\caption{Raman spectra of pristine (up) and saturation potassium doped SWCNTs (down).}
\label{SI_Raman_comp}
\end{center}
\end{figure}

Fig. \ref{SI_Raman_comp}. shows the Raman spectra of both the pristine and saturation K doped materials. The G-mode range indicates a dramatic change in the spectra, as discussed below. In accordance with Ref. \cite{RaoCNTRamanScience}, several softened modes appear in the \emph{mid-frequency} region ($900-1200$ cm$^{-1}$). In Ref. \cite{Rao1998}, these modes were found to be related to a breakdown in the pristine nanotube selection rules associated with the doping.

\begin{figure}[h!]
\begin{center}
\includegraphics[width=.83\linewidth]{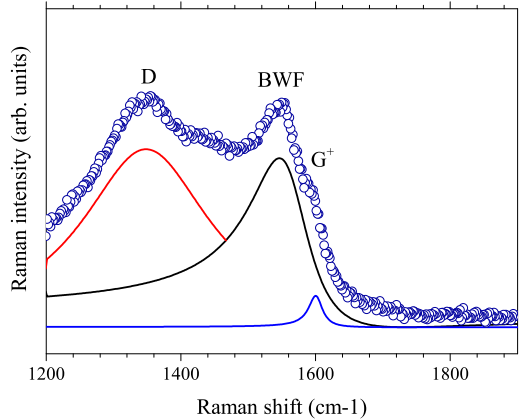}
\caption{Raman spectrum of saturation potassium doped SWCNTs. BWF and G denote the Breit-Wigner-Fano and the graphitic modes, and D represents the defect-related mode.}
\label{SI_Raman_fit}
\end{center}
\end{figure}

In Fig. \ref{SI_Raman_fit}, a fit to the G-mode range Raman spectrum of the saturation doped sample is shown. Three main components are observed. A broad and strong D mode stems from the defects partially created upon alkali doping. The intensity and the significant asymmetry of the Breit-Wigner-Fano (BWF) component points to saturation doping in our sample. The electron transfer to the SWCNTs, measured by the asymmetry of the BWF component is $1/q=-0.305(5)$, a value similar to the one found in Ref. \cite{RaoCNTRamanScience} ($1/q=-0.35$). The G mode at $1600(1)$ cm$^{-1}$, is slightly upshifted compared to the G$^{+}$ mode of the pristine sample ($1592$ cm$^{-1}$, see Fig. \ref{SI_Raman_comp}). In Stage-I graphite (KC$_8$) \cite{JulioPRB2012}, this mode was assigned to regions with lower potassium doping and its intensity was found to decrease compared to the undoped material.

\subsection{Mapping of graphene BZ at KC$_7$ doping}

\begin{figure}[h!]
\begin{center}
\includegraphics[width=.83\linewidth]{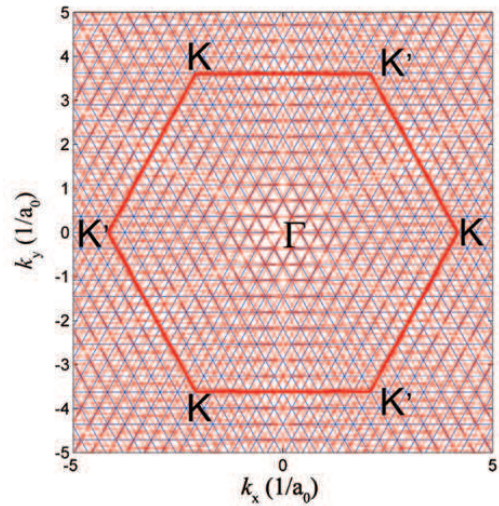}
\caption{Cutting lines shown inside the graphene BZ for the $(10,10)$ and $(11,9)$ nanotubes.}
\label{FigS6_CL_1010}
\end{center}
\end{figure}

The cutting lines for the $(10,10)$ and $(11,9)$ nanotubes is shown in Fig. \ref{FigS6_CL_1010}. When only these two chiralities are considered, a sparsely distributed cutting lines can be recognized. However, our complete calculation is presented in the main text for $81$ different chiralities, which results in a homogeneous coverage of the graphene BZ.

\begin{figure}[h!]
\begin{center}
\includegraphics[width=.83\linewidth]{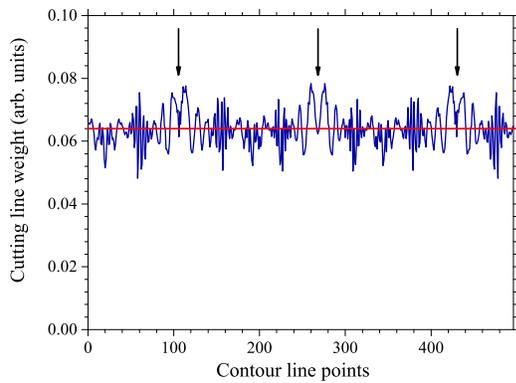}
\caption{Mapping of the graphene BZ by the nanotube ensemble around the contour line defined by the KC$_7$ doping ($E_{\text{F}}=1.73$~eV) as shown in the main text. The arrows indicate the C$_3$ symmetry on the contour line. The solid red line corresponds to the average of the weights around the line.}
\label{SI_weight}
\end{center}
\end{figure}

As pointed out in the main text, the homogeneity of the mapping of the graphene Brillouin zone at a given equi-energetic contour line is of interest for the analysis of our proposed model system. Fig. \ref{SI_weight}. depicts the weight of the coverage of the carbon nanotube ensemble around the equi-energetic contour at the Fermi energy shift corresponding to the KC$_7$ doping. The mapping is found to be homogeneous and it shows small oscillations around the mean value. Interestingly, the trigonal warping does not hinder the C$_3$ symmetry of our illustration. The homogeneity found in this projection is a further proof of the applicability of the ensemble of charge doped SWCNTs for the modeling of biased graphene.

\bibliography{tubes2016july}

\end{document}